\documentclass[10pt,conference]{IEEEtran}
\IEEEoverridecommandlockouts
\usepackage[bottom= 1.15 in,top= 0.7 in,left= 0.57 in,right= 0.68 in]{geometry}
\usepackage{cite}
\usepackage[T1]{fontenc}
\usepackage{aecompl}
\usepackage{fancyhdr}

\usepackage[numbers,sort&compress]{natbib}
\usepackage{amsmath,amssymb,amsfonts}
\usepackage{float} 
\usepackage{booktabs} 
\makeatletter
\renewcommand{\maketag@@@}[1]{\hbox{\m@th\normalsize\normalfont#1}}%
\makeatother
\usepackage{algpseudocode}
\usepackage{graphicx}
\usepackage{caption}
\allowdisplaybreaks[4]
\usepackage{subfigure}
\usepackage{textcomp}
\usepackage{xcolor}
\makeatletter
\newcommand{\removelatexerror}{\let\@latex@error\@gobble}
\makeatother
\usepackage{
    caption} 
\usepackage{float}
\usepackage{graphicx}
\usepackage{textcomp}
\usepackage{xcolor}
\usepackage{bm}
\usepackage{amsmath}
\usepackage{amsthm}
\usepackage{setspace}
\usepackage{algorithm}

\allowdisplaybreaks[4]

\def\BibTeX{{\rm B\kern-.05em{\sc i\kern-.025em b}\kern-.08em
    T\kern-.1667em\lower.7ex\hbox{E}\kern-.125emX}}
\begin{document}

\title{Joint Beamforming and Power Control for D2D-Assisted Integrated Sensing and Communication Networks\\}
\author{Zhenyu~Xue,~\IEEEmembership{}
      Yuang~Chen,~\IEEEmembership{}
      Hancheng~Lu,~\IEEEmembership{}
      Baolin~Chong,~\IEEEmembership{}
      Wanqing~Long~\IEEEmembership{}\\
     University of Science and Technology of China, Hefei, 230027, China\\
     zyxue0919@mail.ustc.edu.cn, yuangchen21@mail.ustc.edu.cn, hclu@ustc.edu.cn, \\ 
     chongbaolin@mail.ustc.edu.cn, madoren@mail.ustc.edu.cn
}
\maketitle

\begin{abstract}
Integrated sensing and communication (ISAC) is an emerging technology in next-generation communication networks. However, the communication performance of the ISAC system may be severely affected by interference from the radar system if the sensing task has demanding performance requirements. In this paper, we exploit device-to-device (D2D) communication to improve communication capacity of the ISAC system. The ISAC system in a single cell D2D assisted-network is investigated, where the base station (BS) performs target sensing and communication with multiple cellular user equipments (CUEs) as well as D2D user equipments (DUEs) simultaneously communicating with other DUEs by multiplexing the same frequency resource. To achieve the optimal communication performance in such D2D-assisted ISAC system, a joint beamforming and power control problem is formulated with the goal to maximize the sum rate while guaranteeing the performance requirements of radar sensing. Due to the non-convexity of the problem, we transform the origin problem into a relaxation form, then, a joint beamforming and D2D power control algorithm is proposed to obtain the solution efficiently. Particularly, a zero-forcing (ZF) beamforming scheme to eliminate the interference from the BS on DUEs is also proposed. Extensive numerical simulations demonstrated that with the assistance of the D2D communications, our proposed algorithm significantly outperforms the baseline schemes in terms of the system sum rate. 
\end{abstract}


\section{Introduction}
With the emergence of an increasing number of applications, various scenarios including smart factory, remote sensing, and vehicle-to-everything not only demand high-speed and low-latency communication services but also require high-precision sensing tasks. However, these complex tasks not only consume more bandwidth leading to spectrum congestion, but also necessitate a more sophisticated hardware platform for implementation. Therefore, as one of the most impacting technologies, integrated sensing and communication (ISAC), has sparked widespread interest in both academia and industry. By sharing the same wireless frequency resource and hardware platform, it can simultaneously achieve both high-speed communication services and high-precision sensing tasks and thus realize system performance gains \cite{liu2022survey}-\cite{liu2022integrated}.  \\
\indent In the ISAC system, the base station (BS) plays a critical role in simultaneously providing radar sensing and communication service, and there has been extensive research on the optimization strategy from the BS perspective in the communication-centric scenario, which aims to achieve better communication performance while satisfying the demand of sensing tasks. The authors in \cite{chen2022generalized} considered the communication-centric scenario where the downlink users are equipped with multiple antennas thereby the system capacity obtains higher channel gain. To improve the spectral efficiency, the ISAC system was \cite{he2023full} investigated in full-duplex work mode, where the BS executes the sensing task and communicates with multiple downlink users and uplink users. The authors in  \cite{zou2022energy} aimed at achieving energy efficient communication performance while guaranteeing target estimation performance. To realize  efficient downlink communication, the authors in \cite{wang2023transmit} proposed to maximize the achievable rate with I/Q-imbalance while satisfying the radar demand constraint.  \\
\indent Most aforementioned work focused on the optimizing resource allocation at the BS side to achieve superior communication performance while satisfying the sensing demand. However, due to the constrained hardware and frequency resource at the BS and the inherent trade-off between sensing and communication performance, the increasingly stringent sensing tasks exacerbates the challenge of obtaining higher communication performance. In this paper, we leverage the device-to-device (D2D) communication into the ISAC system to enhance the system communication capacity. D2D communication  is an effective collaborative communication approach which enables direct communication between adjacent devices without involving the BS. By sharing the spectrum of cellular communication, D2D communication can obviously improve the system spectral efficiency and communication performance \cite{gismalla2022survey}.\\
\indent To exploit the advantages of D2D communication, we propose an ISAC downlink system in a single cell D2D-assisted network, where the BS simultaneously communicates with the cellular user equipments (CUEs) and tracks the target, meanwhile, the D2D user equipments (DUEs) transfer the signal to the other corresponding DUEs. A sum rate maximization problem for the CUEs and the DUEs is formulated while satisfying the radar performance threshold and the power consumption constraint. After relaxing the problem into the convex form, we propose a joint beamforming and D2D power control algorithm to achieve the solution efficiently. A zero-forcing (ZF) beamforming algorithm is also proposed to mitigate the interference from the BS on DUEs. Extensive numerical results demonstrate that our proposed algorithm can converge quickly and achieves significant communication performance gains while satisfying the sensing task requirements, compared to the baseline schemes.\\
\indent The rest of this paper is organized as follows. The system model is presented in Section II. The algorithm under the proposed scheme and the ZF beamforming scheme is described in Section III. Simulation results are shown in Section IV. Conclusion is drawn in Section V.
\section{System Model and Problem Formulation}
\indent We consider a single cellular network with a BS equipped with $N_t$ transfer antennas and $N_r$ receive antennas, $K$ single antenna CUEs and $D$ D2D pairs where each D2D pair contains one single antenna D2D transmitter and one single antenna D2D receiver. As the Fig. 1 depicted, for the downlink transmission of the system, the integrated signal from the BS is simultaneously sent for target sensing and multiple downlink CUEs. Following \cite{he2023full}, \cite{hua2023optimal}, we model the BS downlink signal as \\
\begin{equation}
\setlength\abovedisplayskip{5pt}
\setlength\belowdisplayskip{5pt}
\textbf{x} = \sum_{k=1}^K\textbf{w}_ks_k+\textbf{s}_0,
\label{1} 
\end{equation}

\begin{figure}[t]
	\centering
	\includegraphics[height=7.5cm]{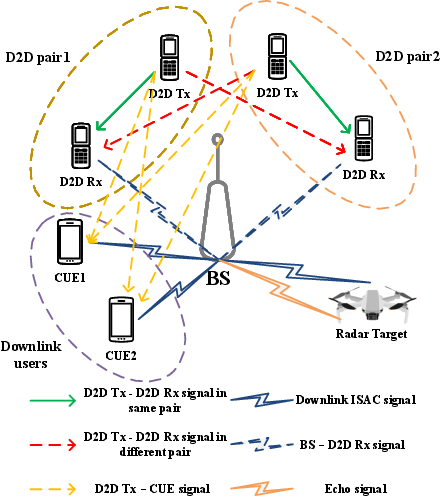}
	\makeatletter\def\@captype{figure}\makeatother
	\setlength{\abovecaptionskip}{-0.01cm}
	\centering
	\caption{ System simultaneously detects the target and downlink communicates while DUEs communicate with other DUEs. }
	\label{fig3}
\end{figure}
\newcounter{sd3}
\setlength{\topmargin}{-0.6 in}
\noindent where $\textbf{w}_k \in \mathcal{C}^{N_t \times 1}$ stands for the beamforming vector associated with the $k^{th}$ CUE, $k \in \{1,...,K\}$, and ${s_k \in \mathcal{C}}$ is the data symbol of the $k^{th}$ CUE with unit power, i.e.,  $E\{|s_k|^2\} =  1$. Besides, $\textbf{s}_0$ is a dedicated radar signal with covariance matrix ${\textbf{W}_0 = E\{\textbf{s}_0\textbf{s}_0^H\}}$. Moreover, we consider a total transmit power constraint for the BS as $\sum_{k=1}^K \|\textbf{w}_k\|^2 + tr(\textbf{W}_0) \leq P^{BS}_{max}$. And the data
symbols of different CUEs and radar signal are assumed to be independent. Except for the BS downlink signal, we also formulate the $d^{th}$ D2D transmitter downlink signal, $d \in \{1,...,D\}$, as \cite{amin2015power}, \cite{mirza2018joint}.\\
\begin{equation}
\setlength\abovedisplayskip{5pt}
\setlength\belowdisplayskip{5pt}
x_d = \sqrt{p_d}s_d,
\label{2} 
\end{equation}
where  ${s_d \in \mathcal{C}}$ is the data symbol of the $d^{th}$ D2D transmitter with unit power, i.e.,  $E\{|s_d|^2\} =  1$. And all DUEs symbols are indenpent. The received signal at the $k^{th}$ CUE and the  $d^{th}$ D2D receiver are given by
\begin{align}
y_k^{CUE} &= \textbf{h}_k^Hx + \sum_{d=1}^D\sqrt{p_d}g_{d,k}s_d + n_k \nonumber \\
 &= \underbrace{\textbf{h}_k^H\textbf{w}_ks_k}_{\text{Desired signal}} + \underbrace{\textbf{h}_k^H\sum_{\substack{k'=1 \\ k' \neq k}}^K\textbf{w}_{k'}s_{k'}}_{\text{Interference signal for other CUEs}} + \underbrace{\textbf{h}_k^Hs_0}_{\text{Sensing signal}} \nonumber \\
&\;\;\;\; + \underbrace{\sum_{d=1}^D\sqrt{p_d}g_{d,k}s_d}_{\text{Interference signal from D2D transmitters}} +  \;\;n_k,
\end{align}
and \\
\begin{eqnarray} 
\label{eq}
 y_d^{D2D} &=& \underbrace{\sqrt{p_d}\rho_{d,d}s_d}_{\text{Desired signal}} + \underbrace{\sum_{\substack{d'=1 \\ d' \neq d}}^D\sqrt{p_{d'}}\rho_{d',d}s_{d'}}_{\text{Interference signal from other D2D transmitters}} \nonumber \\
 &+& \underbrace{\textbf{f}_d^H \sum_{k=1}^K\textbf{w}_ks_k}_{\text{Interference signal from the BS}} + \underbrace{\textbf{f}_d^Hs_0}_{\text{Sensing signal}} + \;\;z_d,
\end{eqnarray} 
\indent Without loss of generality, we assume that the channel gain between any two nodes in the cellular is modeled as a Rayleigh fading channel with the path loss, the channel gain element between node $i$ and node $j$ can be written as \cite{wei2013device} 
\begin{eqnarray} 
\label{eq}
h_{i,j} = c_{i,j}^{-q}h_0,
\end{eqnarray} 
where $c_{i,j}$ is the distance between two node, $q$ is the path loss exponent and $h_0$ represents the small-scale fading which follows complex Gaussian random variable $  \mathcal{C}\mathcal{N}(0,1) $. In (3), $\textbf{h}_k \in \mathcal{C}^{N_t \times 1}$ is the channel gain vector from the BS to the $k^{th}$ CUE, and $g_{d,k} \in \mathcal{C} $ is the channel gain from the $d^{th}$ D2D transmitter to the $k^{th}$ CUE. In (4), $\rho_{d,d} \in \mathcal{C} $ is the channel gain from the $d^{th}$ D2D transmitter to the $d^{th}$ D2D receiver, and $\textbf{f}_d  \in \mathcal{C}^{N_t \times 1}$ is the channel gain from the BS to the $d^{th}$ D2D receiver, $n_k$ and $z_d$ are additive white Gaussian noise $ \mathcal{C}\mathcal{N}(0,N_c) $. Thus, we could derived the SINR as (6), (7) \\
\begin{equation}
\gamma_k^{CUE} = \frac{{|\textbf{h}_k^H\textbf{w}_k|}^2}{\sum_{\substack{k'=1 \\ k' \neq k}}^K|\textbf{h}_k^H\textbf{w}_{k'}|^2 + h_k^H\textbf{W}_0h_k + \sum_{d=1}^Dp_d|g_{d,k}|^2 + N_c },
\label{5} 
\end{equation}
and \\
\begin{equation}
\gamma_d^{D2D} = \frac{p_d |\rho_{d,d}|^2}{\sum_{\substack{d'=1 \\ d' \neq d}}^Dp_{d'}|\rho_{d',d}|^2+ \sum_{k=1}^K|\textbf{f}_d^H\textbf{w}_k|^2 +\textbf{f}_d^H\textbf{W}_0\textbf{f}_d + N_c}.
\label{6} 
\end{equation}
\indent For the sensing procedure, the BS receives the radar reflection signals\cite{lu2022degrees}, we model the echo signals as
\begin{eqnarray} 
\label{eq}
y^{rad} &=& \alpha_0\textbf{a}_r(\theta_0)\textbf{a}_t^T(\theta_0) \textbf{x} + \textbf{c}+ \textbf{z}_s \nonumber \\
&=& \alpha_0\textbf{A}(\theta_0) \textbf{x} + \textbf{c}+ \textbf{z}_s,
\end{eqnarray}
where $\alpha_0$ denotes the path loss and radar cross section,  $\textbf{a}_r(\theta_0) \in \mathcal{C}^{N_r \times 1}$ and $\textbf{a}_t(\theta_0) \in \mathcal{C}^{N_t \times 1}$ represent the steering vector of receive antennas and transfer antennas, which are given as (9) and (10). Besides, $\textbf{c} \in \mathcal{C}^{N_r \times 1}$ represents the undesired reflection signal from the $I$ clutters, which can be modeled as $\textbf{c} = \sum_{i=1}^I \alpha_i\textbf{a}_r(\theta_i)\textbf{a}_t^T(\theta_i) \textbf{x}$, $\textbf{z}_s \in \mathcal{C}^{N_r \times 1}$ represents additive white Gaussian noise with covariance  $N_s\textbf{I}_{N_r}$.
\begin{equation}
\textbf{a}_t(\theta_i) = [1, ... , e^{-j\pi(N_t-1)sin\theta_i}]^T,
\end{equation}
\begin{equation}
\textbf{a}_r(\theta_i) = [1, ... , e^{-j\pi(N_r-1)sin\theta_i}]^T.
\end{equation}
\indent Therefore, we apply a receive beamforming vector $\textbf{t} \in \mathcal{C}^{N_r \times 1}$ to get the desired reflection radar signal and suppress the impact of the communication signal and clutter reflection signal and environment noise. Then, we could formulate the radar signal-clutter-noise ratio (SCNR) as (11)
\begin{eqnarray} 
\label{eq}
\gamma^{rad} &=& \frac{E\{|\textbf{t}^H \alpha_0\textbf{A}(\theta_0) \textbf{x}|^2\}}{E\{|\textbf{t}^H\textbf{c}|^2\}+E\{|\textbf{t}^H\textbf{z}_s|^2\}} \nonumber \\
&=& \frac{|\alpha_0|^2\textbf{t}^H\textbf{A}(\theta_0)\textbf{W}\textbf{A}^H(\theta_0)\textbf{t}}{\textbf{t}^H(\textbf{D}\textbf{W}\textbf{D}^H+N_s\textbf{I}_{N_r})\textbf{t}},
\end{eqnarray}
where $\textbf{D} = \sum_{i=1}^I\alpha_i\textbf{A}(\theta_i)$, and 
\begin{eqnarray} 
\label{eq}
\textbf{W} = E\{\textbf{x}\textbf{x}^H\} = \sum_{k=1}^K\textbf{w}_k\textbf{w}_k^H + \textbf{W}_0,
\end{eqnarray}
denotes the covariance matrix of the downlink ISAC signals. With the discussion above, we aim at maximizing the SCNR, and we note that the maximization problem of the radar SCNR can be seen as the classical minimum
variance distortionless response (MVDR) problem \cite{su2022secure}. Thus, the optimal $\textbf{t}$ can be calculated as 
\begin{eqnarray} 
\label{eq}
\textbf{t}^* &=& argmax\frac{\textbf{t}^H\textbf{A}(\theta_0)\textbf{W}\textbf{A}(\theta_0)^H\textbf{t}}{\textbf{t}^H(\textbf{D}\textbf{W}\textbf{D}^H+N_s\textbf{I}_{N_r})\textbf{t}} \nonumber \\
&=& (\textbf{D}\textbf{W}\textbf{D}^H + N_s\textbf{I}_{N_r})^{-1}\textbf{A}(\theta_0)\textbf{x}.
\end{eqnarray}
\indent Thus, we take the optimal $\textbf{t}^*$ into (11) and the formula will be reformed as
\begin{gather} 
\label{eq}
\gamma^{rad}  =  E\{|\alpha_0|^2\textbf{x}^H\textbf{A}^H(\theta_0)(\textbf{D}\textbf{W}\textbf{D}^H+N_s\textbf{I}_{N_r})^{-1}\textbf{A}(\theta_0)\textbf{x}\} \nonumber \\
= \;\;\;tr\{|\alpha_0|^2\textbf{A}^H(\theta_0)(\textbf{D}\textbf{W}\textbf{D}^H+N_s\textbf{I}_{N_r})^{-1}\textbf{A}(\theta_0) \times  \nonumber \\
 (\sum_{k=1}^K\textbf{w}_k\textbf{w}_k^H+\textbf{W}_0)\}.
\end{gather}
\indent Owing to the complexity of the SCNR of radar, we follow the work \cite{chen2022generalized}, \cite{su2022secure}, by ignoring the dependence of $\textbf{W}$ with respect to $\textbf{x}$. Specifically, in each iteration, we initialize $\textbf{W}$ with the $\textbf{x}$ which is computed in the previous iteration, and $\textbf{x}$ is optimized with the $\textbf{W}$ and $\textbf{W}$ is iteratively updated with the updated $\textbf{x}$ and repeat the computing procedure until convergence. For the communication-centric design, we aim at maximizing the sum of CUEs' and DUEs' achievable rates with satisfying radar performance and power consumption constraints, which can be formulated.
\begin{subequations}
\begin{gather}
\max\limits_{\textbf{w}_k,p_d \geq 0} \sum_{k=1}^Klog_2(1+\gamma_k^{CUE})+ \sum_{d=1}^Dlog_2(1+\gamma_d^{D2D}), \\
s.t. \;\; \gamma^{rad} \geq \eta^{rad},  \\
\sum_{k=1}^K\|\textbf{w}_k\|^2 + tr(\textbf{W}_0) \leq P^{BS}_{max},  \\
p_d \leq P_d , \;\forall d. 
 \end{gather}
\end{subequations}
\indent In the following, we will discuss the solution of (15).
\section{Proposed Algorithm}
\indent In this section, for the reason that the joint design objective function (15) is non-convex both for beamforming downlink vector and D2D power control coefficients. We adopt the semi-definite relaxation (SDR) and successive convex approximation (SCA) method to reform the origin problem to a convex form for the solution. In addition, we adopt the ZF beamforming scheme to cancel out the interference from the BS to the DUEs to acquire the solution. \\
\indent Firstly, we formulate the $\textbf{W}_k = \textbf{w}_k\textbf{w}_k^H, \textbf{W}_k \succeq 0$, following \cite{he2023full}, \cite{ashraf2023joint}, thus, (15) can be reformed as
\begin{subequations}
\begin{gather}
 \max_{\substack{\textbf{W}_k \succeq 0 \\ p_d \geq 0}} \sum_{k=1}^Klog_2(1+\frac{\textbf{h}_k^H\textbf{W}_k\textbf{h}_k}{\sum_{\substack{k'=0 \\ k' \neq k}}^K\textbf{h}_k^H\textbf{W}_{k'}\textbf{h}_k +\sum_{d=1}^Dp_d|g_{d,k}|^2+N_c})   \nonumber \\
+ \sum_{d=1}^Dlog_2(1+\frac{p_d |\rho_{d,d}|^2}{\sum_{\substack{d'=1 \\ d' \neq d}}^Dp_{d'}|\rho_{d',d}|^2+ \sum_{k=0}^K\textbf{f}_d^H\textbf{W}_k\textbf{f}_d + N_c}),   \\
 s.t.  \;\;\gamma^{rad} \geq \eta^{rad}, \\
 \;\sum_{k=0}^Ktr(\textbf{W}_k) \leq P^{BS}_{max},   \\
 p_d \leq P_d , \;\forall d.  
\end{gather}
\end{subequations}
\indent In (16), we drop the rank constraint $rank(\textbf{W}_k) \leq 1, k \in \{0,...,K\}$ to obtain a relaxed form problem following
SDR. Then, the objective function in (16a) can be rewritten as 
\begin{eqnarray} 
\label{eq} (16a)&=&\sum_{k=1}^Klog_2(tr(\textbf{h}_k\textbf{h}_k^H\sum_{k'=0}^K\textbf{W}_{k'})+\sum_{d=1}^D|g_{d,k}|^2p_d+N_c) \nonumber \\
 &-&\sum_{k=1}^Klog_2(tr(\textbf{h}_k\textbf{h}_k^H\sum_{\substack{k'=0 \\ k' \neq k}}^K\textbf{W}_{k'})+\sum_{d=1}^D|g_{d,k}|^2p_d+N_c) \nonumber \\
 &+&\sum_{d=1}^Dlog_2(\sum_{\substack{d'=1 }}^Dp_{d'}|\rho_{d',d}|^2+ \sum_{k=0}^Ktr(\textbf{f}_d\textbf{f}_d^H\textbf{W}_k) + N_c) \nonumber \\
  &-&\sum_{d=1}^Dlog_2(\sum_{\substack{d'=1 \\ d' \neq d}}^Dp_{d'}|\rho_{d',d}|^2+ \sum_{k=0}^Ktr(\textbf{f}_d\textbf{f}_d^H\textbf{W}_k) + N_c).  \nonumber
\end{eqnarray}
\indent From above, the four logarithm functions are concave with respect to $\textbf{W}_k$ and $p_d$, and it is difficult to handle directly. To solve the problem, we adopt SCA method to relax the original problem into a concave form. Firstly, we approximate the second term in (16a) as linear form by  exploiting the first-order Taylor expansion, it holds that
\begin{eqnarray} 
\label{eq}
&\;&log_2(tr(\textbf{h}_k\textbf{h}_k^H\sum_{\substack{k'=0 \\ k' \neq k}}^K\textbf{W}_{k'})+\sum_{d=1}^D|g_{d,k}|^2p_d+N_c)  \nonumber \\
&\leq& log_2(tr(\textbf{h}_k\textbf{h}_k^H\sum_{\substack{k'=0 \\ k' \neq k}}^K\textbf{W}_{k'}^{(i-1)})+\sum_{d=1}^D|g_{d,k}|^2p_d^{(i-1)}+N_c)  \nonumber \\
&\;&+\;\;\;\beta_k^{(i-1)} \times \frac{tr(\textbf{h}_k\textbf{h}_k^H\sum_{\substack{k'=0 \\ k' \neq k}}^K(\textbf{W}_{k'}-\textbf{W}_{k'}^{(i-1)}))}{ln2} \nonumber \\ 
&\;&+\;\;\; \beta_k^{(i-1)} \times\frac{\sum_{d=1}^D|g_{d,k}|^2(p_d-p_d^{(i-1)})}{ln2} \nonumber \\ 
&\triangleq& \varphi^{CUE}_k,
\end{eqnarray}
where $\beta_k^{(i-1)} = 1/{(tr(\textbf{h}_k\textbf{h}_k^H\sum_{\substack{k'=0 \\ k' \neq k}}^K\textbf{W}_{k'}^{(i-1)})+\sum_{d=1}^D|g_{d,k}|^2} \times $ \\ $p_d^{(i-1)} +N_c)$, and $\textbf{W}_{k}^{(i-1)}$, $p_d^{(i-1)}$ are the solution to $\textbf{W}_{k}$, $p_d$ obtained in the $(i-1)$-th iteration. For the last term in (16a), we use the same approximation method in (17) with handling the dual problem. Therefore, the last term in (16a) can be relaxed as
\begin{eqnarray} 
\label{eq}
&\;&log_2(\sum_{\substack{d'=1 \\ d' \neq d}}^Dp_{d'}|\rho_{d',d}|^2+ \sum_{k=0}^Ktr(\textbf{f}_d\textbf{f}_d^H\textbf{W}_k) + N_c) \nonumber \\
&\leq& log_2(\sum_{\substack{d'=1 \\ d' \neq d}}^Dp_{d'}^{(i-1)}|\rho_{d',d}|^2+ \sum_{k=0}^Ktr(\textbf{f}_d\textbf{f}_d^H\textbf{W}_k^{(i-1)}) + N_c)  \nonumber \\
&\;&+\;\;\;\delta_d^{(i-1)} \times \frac{\sum_{d=1}^D|\rho_{d',d}|^2(p_d-p_d^{(i-1)})}{ln2} \nonumber \\ 
&\;& +\;\;\;\delta_d^{(i-1)} \times \frac{\sum_{k=0}^Ktr(\textbf{f}_d\textbf{f}_d^H(\textbf{W}_k-\textbf{W}_k^{(i-1)}))}{ln2} \nonumber \\ 
 \nonumber \\ 
&\triangleq& \phi^{D2D}_d, 
\end{eqnarray}
in (18), where $\delta_d^{(i-1)} = 1/{(tr(\textbf{f}_d\textbf{f}_d^H\sum_{\substack{k=0}}^K\textbf{W}_{k}^{(i-1)})})+\sum_{\substack{d'=1 \\ d' \neq d}}^D|\rho_{d',d}|^2 \times p_{d'}^{(i-1)}+N_c)$. By introducing the relaxation term (17), (18), the origin problem can be reformed as (19) and the relaxed problem can be easily solved owing to the concavity. 
\label{eq}
\begin{eqnarray}
&\max\limits_{\textbf{W}_k,p_d}& \sum_{k=1}^Klog_2(tr(\textbf{h}_k\textbf{h}_k^H\sum_{k'=0}^K\textbf{W}_{k'})+\sum_{d=1}^D|g_{d,k}|^2p_d+N_c) \nonumber \\ 
&+& \sum_{d=1}^Dlog_2(\sum_{\substack{d'=1 }}^Dp_{d'}|\rho_{d',d}|^2+ \sum_{k=0}^Ktr(\textbf{f}_d\textbf{f}_d^H\textbf{W}_k) + N_c) \nonumber \\
&-& \sum_{k=1}^K\varphi^{CUE}_k -\sum_{d=1}^D\phi^{D2D}_d, \nonumber \\  
& s.t.& (16b),(16c),(16d).
\end{eqnarray}
\indent In addition to the scheme of beamforming matrix as we stated above, we also give the fixed beamfoming scheme based on the ZF method for performance comparison where the BS will downlink its data in the null space of the channels between the BS and the D2D receivers, which can effectively cancel out the interference from the BS downlink signal to the DUEs. Specifically, the downlink beamforming covariance matrix should be satisfied with
\label{eq}
\begin{eqnarray}
\textbf{f}_d^H\textbf{W}_k\textbf{f}_d = 0 , \;\forall d, \;\forall k.
\end{eqnarray}
\indent By introducing (20), the origin problem (16) can be reformed as (21) and it can be seen that the interference from BS to D2D user is cancelled out so that to obtain the better D2D communication performance.
\begin{subequations}
\begin{gather}
\max_{\substack{\textbf{W}_k \succeq 0 \\ p_d \geq 0}}  \sum_{k=1}^Klog_2(1+\frac{\textbf{h}_k^H\textbf{W}_k\textbf{h}_k}{\sum_{\substack{k'=0 \\ k' \neq k}}^K\textbf{h}_k^H\textbf{W}_{k'}\textbf{h}_k   +\sum_{d=1}^Dp_d|g_{d,k}|^2+N_c})   \nonumber \\
 +\sum_{d=1}^Dlog_2(1+\frac{p_d |\rho_{d,d}|^2}{\sum_{\substack{d'=1 \\ d' \neq d}}^Dp_{d'}|\rho_{d',d}|^2+ N_c}),    \\
s.t. \;\;\;\;\textbf{f}_d^H\textbf{W}_k\textbf{f}_d = 0 , \;\forall d, \;\forall k,\\
(16b),(16c),(16d).
\end{gather}
\end{subequations}
\indent It is clear to see (21b) is linear form constraint condition, therefore the problem is similar to (19) and it can be solved by the same algorithm given in (17), (18) to relax the problem into a convex form. We summarize the algorithm in the proposed scheme and the ZF scheme as Algorithm 1 in the following. 
\begin{algorithm}
	\renewcommand{\algorithmicrequire}{\textbf{Input:}}
	\renewcommand{\algorithmicensure}{\textbf{Output:}}
	\caption{Joint beamforming and D2D power control algorithm}
	\label{alg1}
	\begin{algorithmic}[1]
		\Require Initialization:$\left\{ {\textbf{W}_k^{(0)}} \right\},\left\{ {p_d^{(0)}} \right\},$ iteration index $\emph{i}$ = 0, and maximum number of iteration = $\emph{I}_{max}.$
		\Repeat
            \State Set $\emph{i}$ = $\emph{i}$ + 1;
		\State Update (20) (or (22)) with $\left\{ {\textbf{W}_k^{(\emph{i}-1)}} \right\},\left\{ {p_d^{(\emph{i}-1)}} \right\};$
		\State Solve (20) (or (22)) and update $\left\{ {\textbf{W}_k^{(\emph{i})}} \right\},\left\{ {p_d^{(\emph{i})}} \right\};$
		\Until the objective converges or $\emph{i}$ = $\emph{I}_{max}$.
		\Ensure  $\left\{ {\textbf{W}_k} \right\},\left\{ {p_d} \right\}.$
	\end{algorithmic}  
\end{algorithm}
\vspace{-0.55em}
\section{Simulation results}
\vspace{-0.55em}
\indent In this section, we provide the numerical simulation results based on the algorithm we gived in the proposed scheme and the ZF scheme, and we compare our algorithm performance with different benchmarks under various scenarios. 
\subsection{Simulation Setup}
\indent We assume that the numbers of receive antennas and transfer antennas are $N_t$ = 8, $N_r$ = 8. From the sensing perspective, the target is located at the $\theta_0 = 0$  of the BS, and $I$ = 2 clutters are distributed in the environment which are located at $\theta_1 = -\pi/6$  and $\theta_2= \pi/6$. Besides, we set radar channel power gains of the target and two clutters as $|\alpha_0|^2$/$N_s$ = 20 dB, and $|\alpha_1|^2$/$N_s$ = $|\alpha_2|^2$/$N_s$ = 80 dB, and radar channel noise power $N_s$ = -70 dBm. From the communicating perspective, we assume that  $K$ = 3 downlink users and $D$ = 2 D2D pairs within the communication coverage of the BS, the average distance between the BS and UE is 100 m and the average distance inside the D2D pair is 10 m. The path loss exponent is set to 2. The power budget of the BS $P^{BS}_{max}$ is set to 30 dBm and each D2D pair's power budget is set to 10 dBm, the communicating channel noise power $N_c$ = -70 dBm. And then the maximum number of iterations $\emph{I}_{max}$ = 8.  \\
\indent For verifying the proposed algorithm performance effectiveness, we consider three baseline schemes for comparison.\\
\indent 1) Sensing-only Scheme: In the sensing-only scheme, we only take maximizing the radar SCNR under transmitting power limits into consideration, while ignoring the constraints of communication.\\
\indent 2) Communication-only Scheme: Similar to the sensing-only scheme, this scheme considers maximizing the sum rate of CUEs and DUEs while neglecting the thresholds of radar performance demanded.\\
\indent 3) Fixed D2D Scheme: The scheme aims at maximizing the sum rate of the system while fixing D2D transmitting power and satisfying the radar performance constraints.
\subsection{Numerical results}
\indent Fig. 2 illustrates the convergence performance of system channel capacity including all CUEs and DUEs for our proposed algorithm. The algorithm achieves convergence within a few iterations both in the proposed scheme and the ZF scheme. Moreover, with the comparison of the proposed communication-only scheme and ZF communication-only scheme, the system capacity is lower than the communication-only scheme for the reason that sensing task occupied part of power allocation in the BS. 
\begin{figure}[htbp]
	\centering
	\vspace{-0.7em}
	\includegraphics[height=6cm]{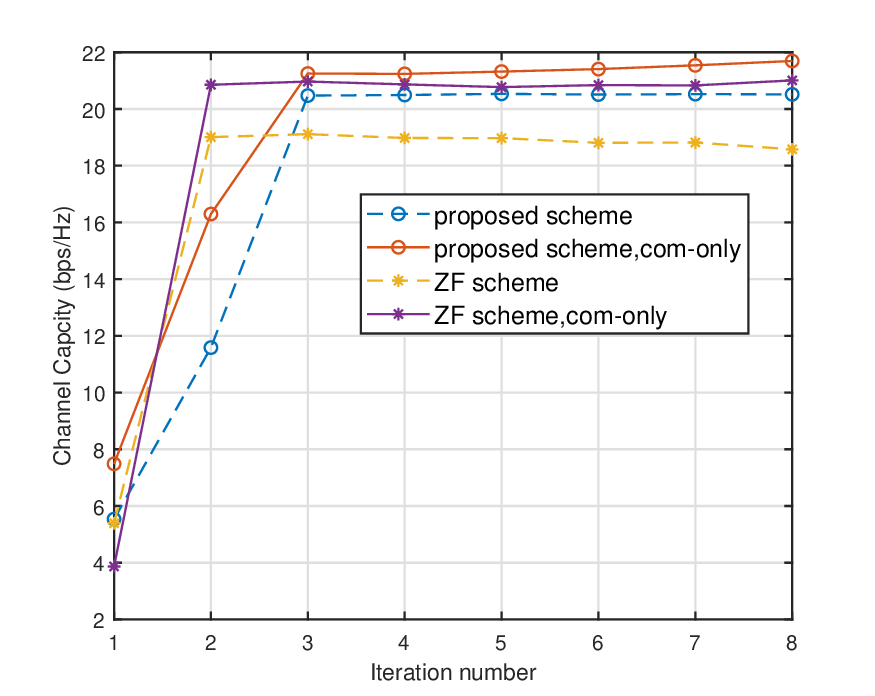}
	\makeatletter\def\@captype{figure}\makeatother
	\setlength{\abovecaptionskip}{-0.01cm}
	\centering
	\caption{ Convergence performance of algorithm of different schemes. }
	\label{fig3}
\end{figure}
\newcounter{sd2}

\indent As Fig. 3 plotted, the proposed beampatterns under different SCNR constraints in the proposed scheme and the ZF scheme with the beampattern under the sensing-only scheme are shown. We firstly introduce the definition of the proposed beampattern based on the optimal receive beamformers $\textbf{t}^*$ as $P(\theta) = |{\textbf{t}^*}^H\textbf{A}(\theta)\textbf{x}|^2$. It can be seen that beampattern in the proposed scheme and the ZF scheme have similar performance, the main beam is located at 0, and the deep nulls are placed at the clutters' angles as we set. Moreover, with the SCNR constraints is becoming larger, the radar performance becomes better and the shape is more closed to the sensing-only scheme beampattern. \\
\begin{figure}[t]
\centering  
\subfigure[Beampattern of the proposed scheme]{
\label{Fig.sub.1}
\includegraphics[height=3.22cm]{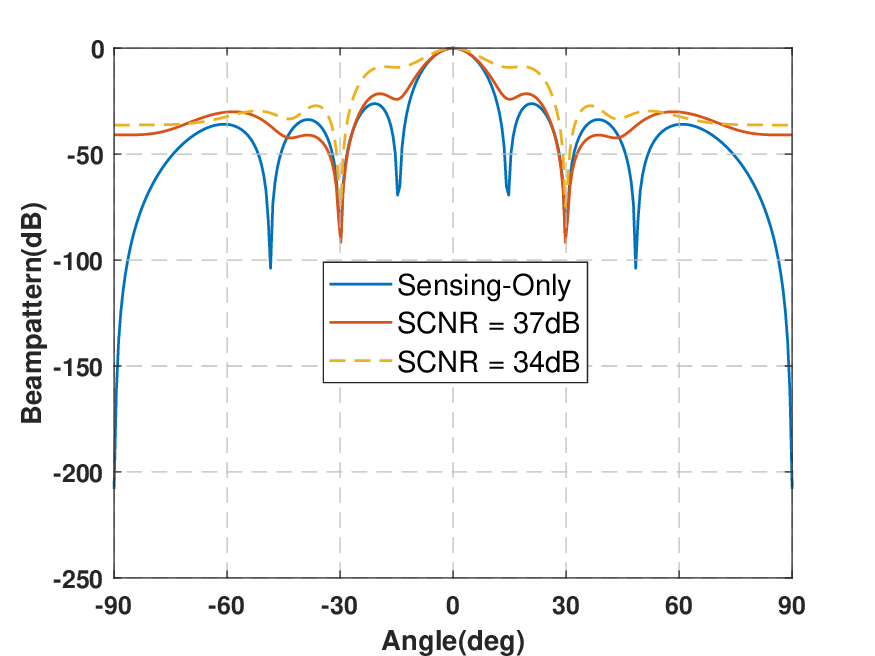}}
\hspace{-0.3cm}
\subfigure[Beampattern of the ZF scheme]{
\label{Fig.sub.2}
\includegraphics[height=3.22cm]{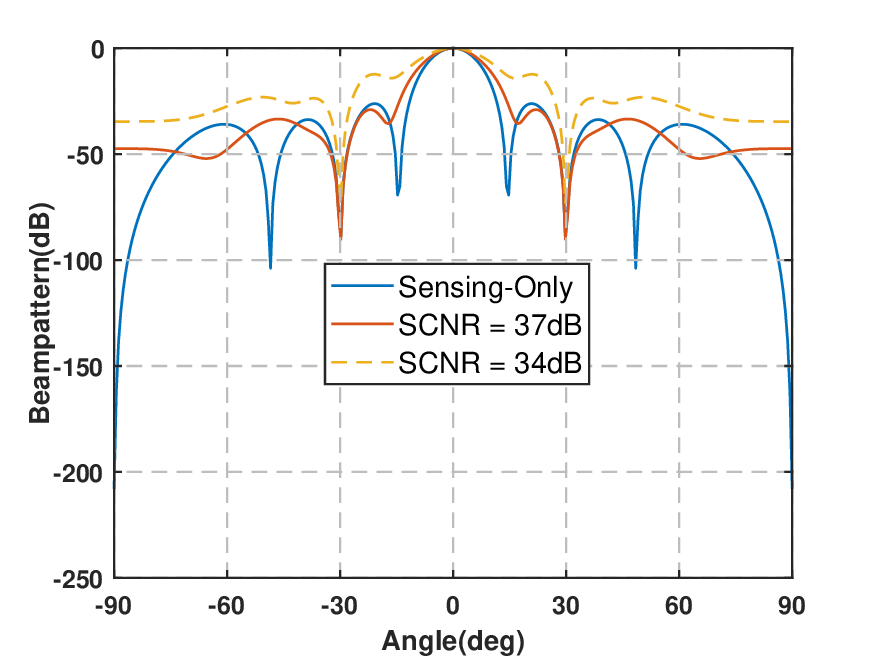}}
\caption{The optimized beampatterns with different SCNR constraints}
\label{1}
\end{figure}
\indent In Fig. 4, we present an analysis of the performance trade-off between the overall communication performance of the system including all the CUEs,  and DUEs versus the system's radar sensing demand. In the case of the communication-only scheme, the system sum rate remains constant, as the radar performance threshold is neglected. From the fixed D2D scheme, in which we fixed the power control and the system communication is only affected by the downlink beamforming matrix, and it can be seen that with the increase of the radar SCNR, the system channel capacity decreases, specifically, when the radar SCNR constraint exceeds 34 dB, for each additional one dB increase, the system's communication performance deteriorates more severely. As a result, when the radar performance requirements become more stringent, the communication performance experiences a more significant decline. For our proposed scheme, with the sensing SCNR constraints become higher, the system achievable rate becomes a little lower, even with the radar constraint being stringent, the system capacity does not severely deteriorate, the reason is that the downlink communication performance is limited due to the radar task, the sum rate of DUEs becomes higher with the downlink communication capacity decreases because the D2D communication and the cellular communication work in the same band, thereby ensuring that the overall system communication capacity remains relatively stable even as the radar performance requirements increase. The ZF scheme exhibits that when the SCNR constraint is identical, the system communication performance is a little worse than the system capacity under the proposed scheme and is higher than the sum rate of the fixed scheme, and the cause is that the proposed scheme joint optimally optimize the downlink beamforming matrix and D2D power control, meanwhile the fixed D2D scheme only optimized the downlink beamforming and the ZF scheme is optimized under the constraint of ensuring the downlink signal transmitting in the null space of the D2D channel. Furthermore, it is evident that the system capacity in the ZF scheme is comparatively smaller than the decrease in communication capacity of the proposed scheme, because the ZF scheme cancels out the interference from the BS to the DUEs, so the D2D communication contributes more channel capacity to the system without downlink interference. From analysis above illuminated that by introducing D2D in the system, the system communication capacity can remain unchanged or be maintained within a relatively small, thus D2D improves the system communication capacity even when the sensing task has a stringent performance demand. 
\section{Conclusion}
\indent In this paper, we investigated the ISAC system with D2D-assisted networks and proposed the joint beamforming and D2D power control strategies. We formulated the maximization problem of the system sum rate  under the radar performance threshold and power consumption constraint. Then we used the relaxation method to convert the non-convex origin problem into the convex form and obtain the solution, besides, we proposed the ZF beamforming scheme to acquire the solution for comparison. Simulation results present that with the same radar performance constraints, our proposed solution improves system communication performance by 22$\%$ compared to the fixed D2D scheme, which illustrates demonstrates that even with stricter radar constraints, the D2D-assisted system's communication performance can remain stable rather than experiencing a sharp decline.

\begin{figure}[htbp]
	\centering
	\vspace{-0.7em}
	\includegraphics[height=6cm]{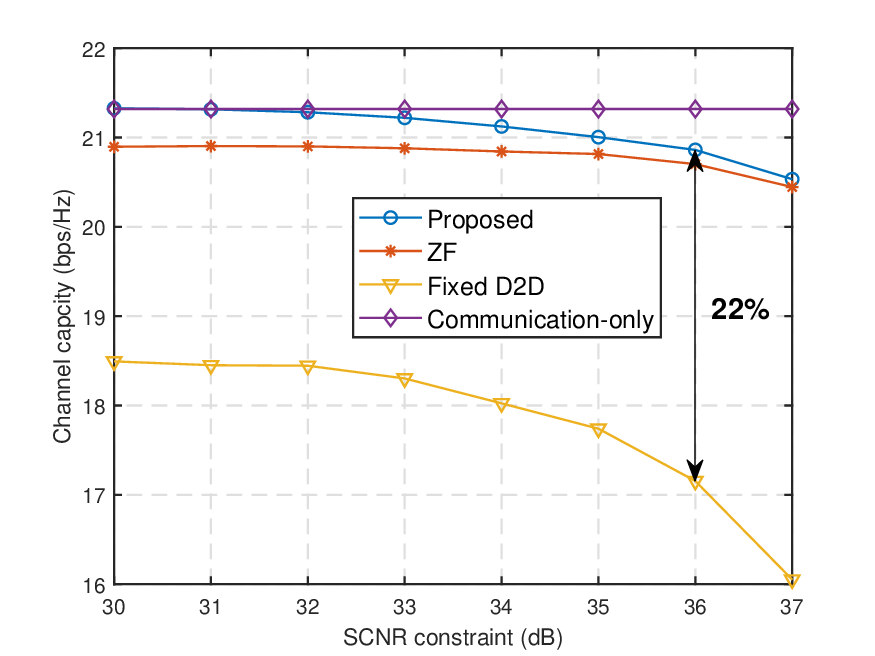}
	\makeatletter\def\@captype{figure}\makeatother
	\setlength{\abovecaptionskip}{-0.01cm}
	\centering
	\caption{ System channel capacity versus SCNR constraints. }
	\label{fig3}
\end{figure}
\newcounter{sd4}
\vspace{12pt}
\footnotesize
\bibliographystyle{IEEEtranN}
\bibliography{IEEEabrv,reference}
\end{document}